# Single image sensorless adaptive optics using phase-diversified multifocal array

Biwei Zhang, Qi Hu, Jingyu Wang and Martin Booth

Department of Engineering Science, University of Oxford, Oxford OX1 3PJ, UK

## Abstract

Image-based sensorless adaptive optics is commonly used to improve microscope image quality through aberration compensation. However, its sequential nature leads to repeated sample exposures and a prolonged aberration estimation and correction process. In this work, we used a phase-diversified foci array to enhance the aberration-related information in multi-photon fluorescence microscope images. We showed that, through use of a neural network-based image processing algorithm, we can extract aberration from a single image frame. We discussed how the number of foci can affect the wavefront correction on a bead sample, and we showed that six foci, in general, perform better than two-foci or ten-foci arrangements. We also applied our technique to correct aberrations in biological samples, including a fixed brain tissue and BPAE cells slide.

## Introduction

Adaptive optics (AO) has become an important technique in microscopy for correcting aberrations introduced by either the optical system or the specimen [1-3]. Among the various implementations, image-based sensorless AO is particularly attractive because it requires no wavefront sensor and is therefore straightforward to integrate into existing microscopes [4]. In its conventional implementation, a sequence of images is acquired while applying different known bias aberrations with an adaptive optical element. The unknown aberration is then estimated from the resulting image set using a suitable mathematical model. In essence, the information about aberrations is encoded in the set of images and the algorithm extracts an estimate of these aberrations. Early sensorless AO methods relied on maximising an image quality metric, such as image brightness [5-9] or sharpness [10,11], to determine the optimal correction. More recently, machine learning approaches have demonstrated substantially improved performance by learning a direct mapping between image data and aberration coefficients [12-15]. When combined with appropriate image pre-processing, these methods have extended the operating range of sensorless AO while reducing the number of images required for reliable aberration estimation In previous work [16], we showed that aberration could be extracted from as few as two biased images, although with reduced accuracy compared with acquisitions using a larger number of bias images.

To further improve sensorless AO method, it is essential to increase the amount of aberration-sensitive information in the acquired images available to the estimator. If one could better condition the encoding of information about aberrations in each acquired image, this would enable accurate estimation with fewer image acquisitions, thereby reducing both acquisition time and specimen exposure. This perspective naturally leads to the question of whether multiple aberration modes can be estimated from a single image by encoding sufficient aberration-sensitive information.

In this work, we demonstrate a sensorless AO framework for two-photon excitation fluorescence microscopy that requires only a single image acquisition for aberration estimation. Rather than acquiring multiple sequential phase-diversified images, we use holographic beam shaping to generate a spatially separated phase-diversified foci array to acquire a single image frame. [Figure 1 (a)] Effectively, we superimpose otherwise sequentially obtained images into one frame with lateral separations. A neural network is then used to estimate the aberration coefficients directly from the resulting single image. [Figure 1 (b)] This approach takes inspiration from the holographic modal wavefront sensor by generating a multi-foci pattern to encode different phase mode information in the collection of foci [17-19]. This original holographic wavefront sensor involved measurements on isolated focal spots and was not compatible with an imaging system. By combining an equivalent multi-foci method with neural network estimation, these advances significantly reduce the acquisition burden of image-based sensorless AO while maintaining accurate aberration estimation.

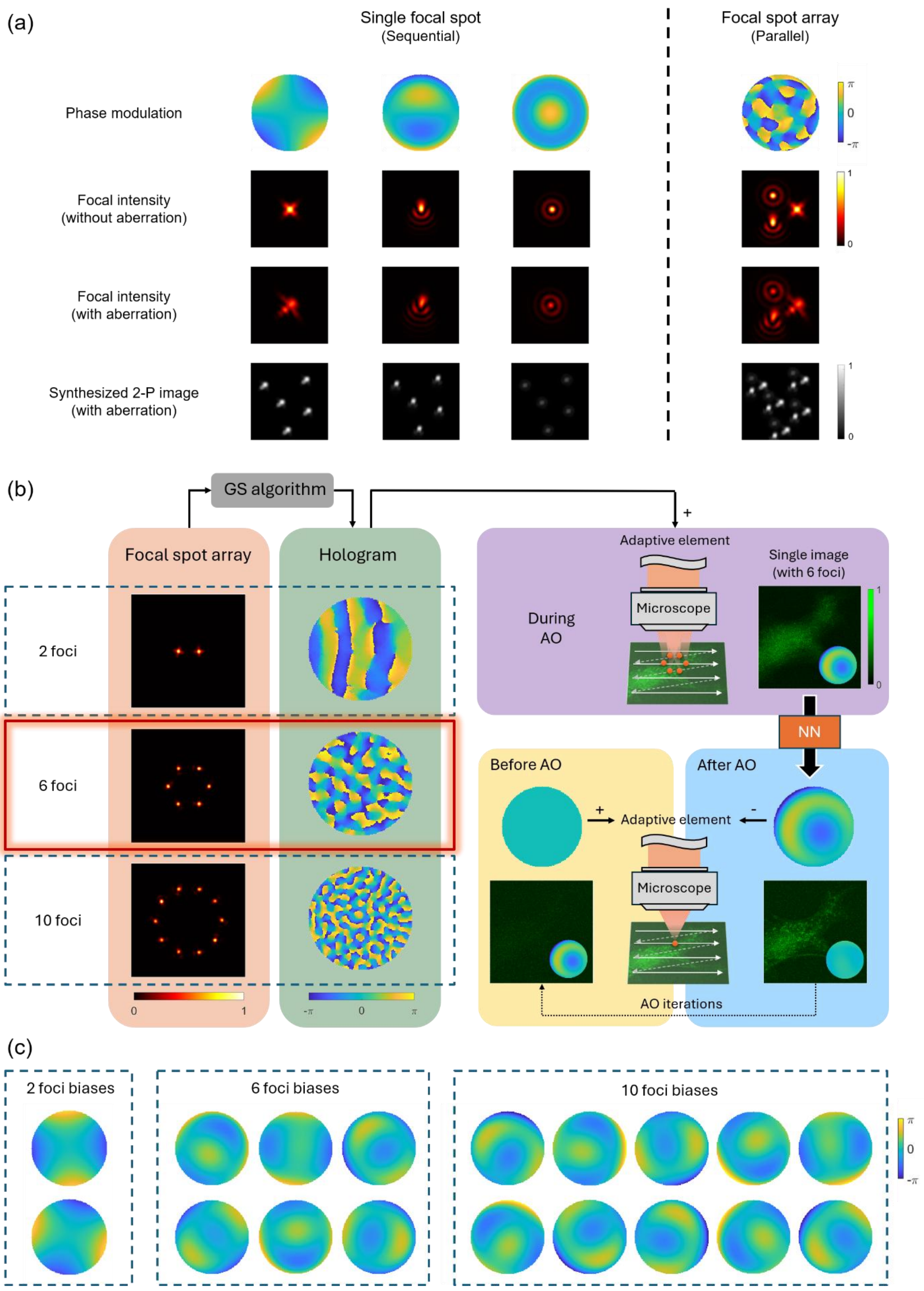

*Figure 1 (a) illustrations to show how different phase modulations and the computer generated hologram can affect the focal spots and the synthesized 2-P images. (b) The sensorless AO framework proposed in this work and how it could be used in a 2-P adaptive microscope. (c) The derived phase biases we used to generate the spatial multiplexed foci array. For the derivation, please see the "Multi-foci generation" subsection.*

## Concept

The concept was demonstrated using a two-photon excitation fluorescence microscope. The conventional approach to sensor-less AO involves using the same adaptive element both to correct aberrations and to introduce bias aberrations in a sequential manner. In this case, a spatial light modulator was chosen as the adaptive element, which was used for aberration correction, but also to display phase holograms to generate an array of biased foci in parallel. The hologram was designed such that each of the foci contained a pre-chosen bias aberration that was different from the others. This multi-foci array was scanned across the specimen in the same manner as typically used for a single focus in scanning microscopy. [Figure 1] The resulting image was, in effect, a superposition of spatially offset biased images, equivalent to those that would have been obtained through sequential acquisition with the same bias aberrations applied. This single image, in principle, contains similar information to the set of separate biased images, albeit with some reduced information content due to the superposition.

The challenge of this method is to extract the required wavefront information from the complex output image, in which it is encoded. The information extraction must also be independent of the unknown specimen structure. This would be an impractical task from the multi-foci image using conventional sensorless AO approaches. However, machine learning methods are ideally suited to the extraction of information from such images.

For this purpose, we used a convolutional neural network to process the multi-foci images and extract phase information in the form of aberration modal coefficients. For generation of training data, we created a digital twin of the two-photon microscope to produce simulated images with randomly introduced aberrations, the designed multi-foci spot pattern, and randomly synthesised sample structures. The resulting dataset was used to train the network, which was tested on experimental images of beads and biological samples to demonstrate its performance.

### Experimental set-up

The experiment was based around a two-photon laser scanning microscope that incorporated a Hamamatsu liquid-crystal spatial light modulator (LCOS-SLM X15213-02) configured for phase-only modulation. The design of the microscope is shown below. The laser source (Newport Spectra-Physics Deep-See) was operated at 850 nm.

To generate the multi-foci array at the focal plane of the objective, a hologram was displayed on the spatial light modulator (SLM), superimposed on a blazed grating, which enabled separation of the modulated light into the first diffraction order, away from the zero-order unmodulated light. A spatial filter (SF) at the Fourier plane of the SLM allowed the first-order foci pattern to pass through and block unwanted light. The two conjugated galvanometer mirrors (X and Y) scanned the foci pattern across the field of view.  The fluorescence signal was separated by a short pass dichroic (cut-off at 700 nm) and an emission filter (cut-off at 568 nm), then detected by the photomultiplier tube (PMT).

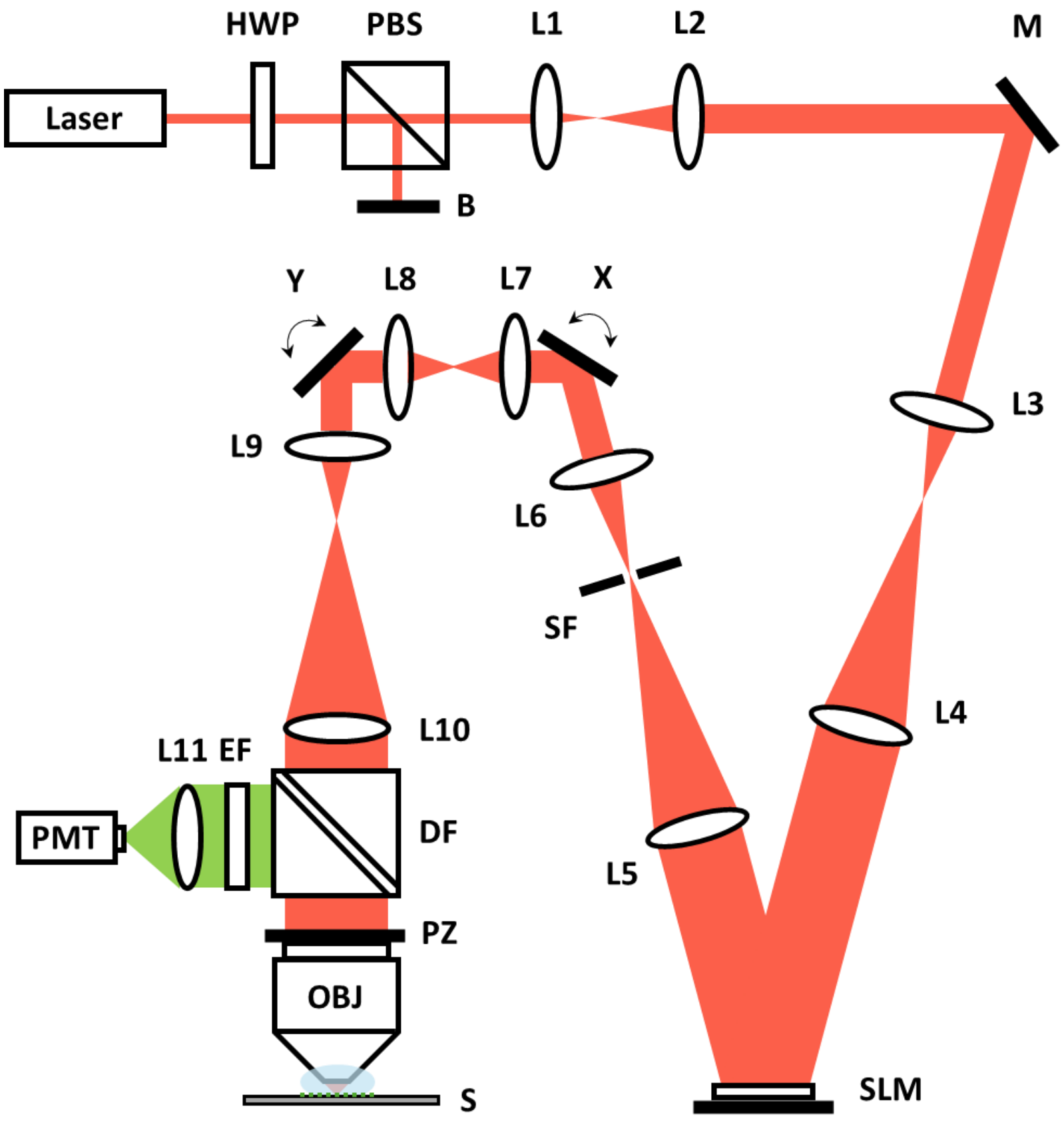

*Figure 2: Schematic of the adaptive two photon microscope. HWP: half waveplate. PBS: polarised beam splitter. L1-L11: tube lenses. M: silver mirror. SLM: Hamamatsu liquid-crystal spatial light modulator. SF: spatial filter. X: x direction galvo mirror. Y: y-direction galvo mirror. DF: dichroic filter. EF: emission filter. PMT: photomultiplier tube. PZ: piezo stage. OBJ: Olympus objective lens NA 1.15 water immersion. S: sample.*

## Multi-foci generation

The hologram to generate the multi-foci pattern was computed through a modified Gerchberg-Saxton (GS) algorithm [20,21]. Due to the focal shape ambiguity of even symmetry Zernike modes, the in-focus focal spot shape is identical for both the positive and negative biases of even Zernike modes. Hence, the GS algorithm used two defocused planes rather than a single in-focus plane for optimisation.

In each case, the desired number of biased foci in the array was chosen. The foci were arranged in a circular shape with equal separation, which was chosen heuristically to be 8 × FWHM of the diffraction-limited 2-P point spread function.  This spacing was a compromise between ensuring that the foci were well separated so did not overlap when aberrated, and occupying a narrow region. This arrangement was chosen for effective demonstration purposes, but the method is not limited to this selected pattern.

We chose three foci arrangements – 2, 6 and 10 foci arrays, for fair comparisons to understand how aberration information could be better encoded in microscope images. [Figure 1(c)] We chose these three representative numbers because they correspond to three typical sensorless AO arrangements. 2 foci was shown to be the minimal number of biases one should use for determining aberrations. 6 foci and 10 foci, in our case, corresponds to N+1 and 2N number of image acquisitions when measuring aberrations consisting of N modes (5 modes in our case).

The bias aberration applied to each of the foci were derived as combinations of the five modes astigmatism ($Z5$ and $Z6$), coma ($Z7$ and $Z8$) and spherical ($Z11$). The specific combinations were generated through optimisation of the aberration-relevant Fisher information content in the output images. More information on how the optimisation was conducted can be found in [22].

### Image processing algorithm

To process the images, we used a convolutional neural network with a ResNet 18 architecture. We modified the network such that the number of outputs matched the number of aberration coefficients involved in the experiments. For most of the work presented below, we included for illustration of the method as specimen aberrations the five Zernike modes astigmatism ($Z5$ and $Z6$), coma ($Z7$ and $Z8$) and spherical ($Z11$).

Using a digital twin model of the two-photon microscope imaging process, we synthesised a training dataset of 100,000 examples and a validation dataset of 1,000 examples for the network training. For each dataset, we introduced a random aberration, which followed a n-sphere uniform-distribution [23] with the root-mean-square value smaller than 2.236 rad. We used an Adam gradient descent optimizer for 50 epochs of training and the loss function was the RMS value of the residual aberration. We used a dynamic learning rate adjustment starting from $10^{-3}$ and decayed by 2/3 each time till $10^{-5}$, triggered when the validation loss failed to drop in two consecutive epochs. The inference time of the trained neural network was around $8\ ms$.

## Results

We systematically analysed sensing performance on a sample consisting of $2\mu m$ diameter Thermo Fisher Scientific FluoSpheres (505/515). The aberration correction accuracy was assessed by understanding the influence of signal level jointly with the number of foci ($k$) used in the pattern. Three separate neural networks were trained, each corresponding to a different foci pattern for $k$=2, 6, and 10, which were tested over a range of different imaging conditions. We adjusted the total laser power by rotating the half waveplate (HWP) such that the total at-sample laser power for single focal spot ($k = 1$) equalled $5.8mW$ (low SNR), $11.6mW$ (medium SNR) and $23.2mW$ (high SNR) respectively. The total at-sample laser power was adjusted for each multi-foci design to keep the overall two-photon fluorescence signal constant for one image frame (assuming a uniform fluorophore distribution in the specimen). This means that for a $k$-foci design, the total at sample laser power was adjusted to $\sqrt{k} \times 5.8mW$ (low SNR), $\sqrt{k} \times 11.6mW$ (medium SNR) and $\sqrt{k} \times 23.2mW$ (high SNR) respectively.

For each SNR condition, each multi-foci pattern and each aberration range, we collected 20 sets of results, and generated statistical results presented in Figure 3. For each set of the results, the aberration was randomly generated with a RMS value within the range labelled in the figure.

We used an intensity metric ($y_I$) as a proxy to quantify the aberration correction performance. $y_I$ was defined as the summation of the highest 350 pixel values and normalised by the value before correction. As we typically used sparse samples, by summing the highest pixel values in this way, this metric represented the variations of signal fluorescence intensity and minimised the effects of noise and background.

The results indicated clearly that the method provided promising performance across a range of scenarios, resulting in brighter images after the correction process. The methods showed strongest correction ability, when the initial aberration was large and when the SNR was high. The pattern with 2-foci was most sensitive to noise, as evidenced at low SNR, and the 10-foci pattern performed the best when there was sufficient fluorescent signal, corresponding to high SNR. Across all the different imaging scenarios, the 6-foci pattern in general provided the most consistent results and was the optimised option compared to 2- and 10-foci designs. In particular, the 6-foci pattern performed well for the lowest SNR demonstrations, which indicates an optimum balance between phase diversity in the foci and having sufficient emission from each focus to provide reasonable signal above the background level.

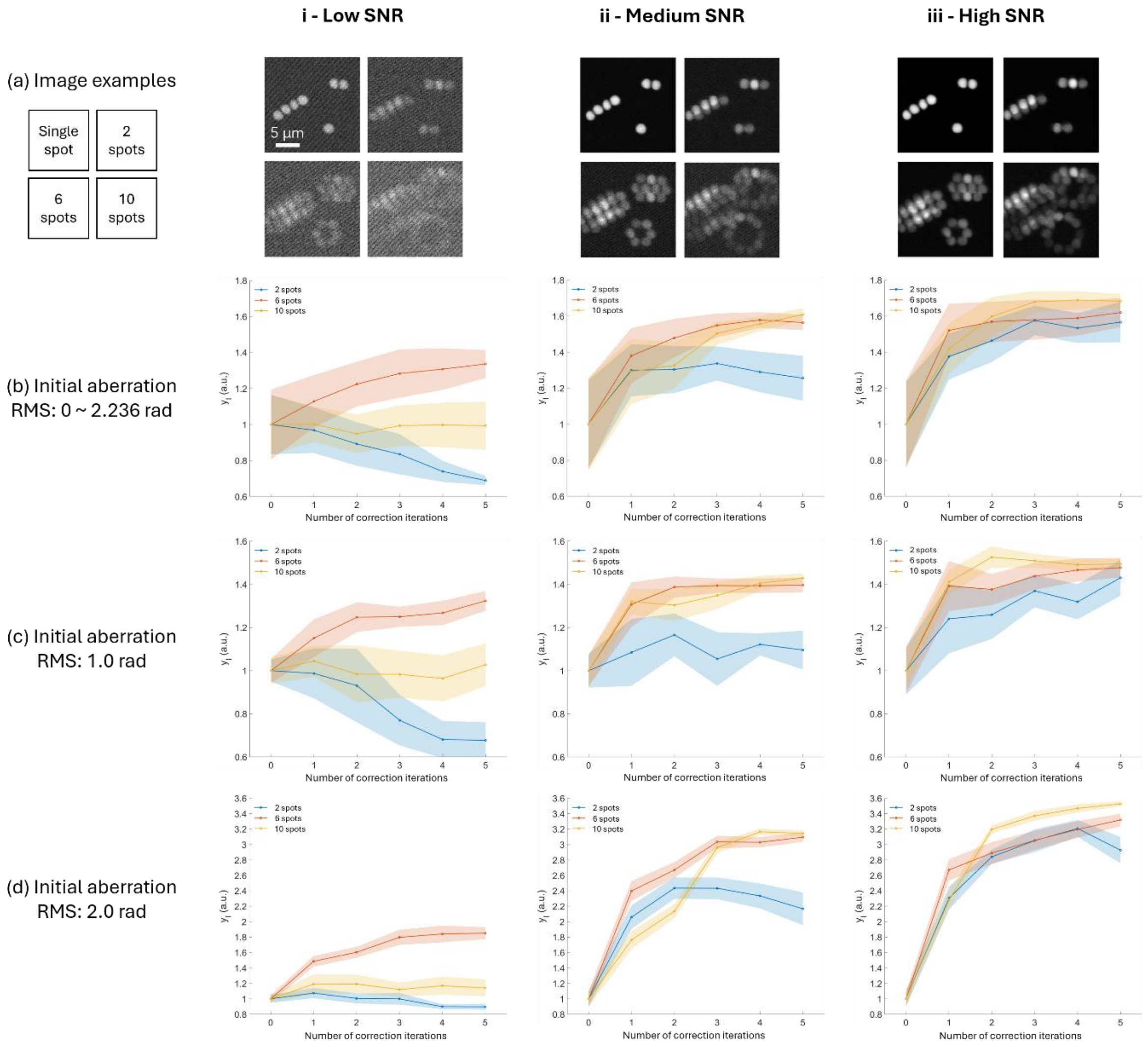

*Figure 3 (a) example images to illustrate the selected SNR levels and how they affected the microscope images. (b-d) Five iterations of aberration corrections with different foci-array at different noise levels when the random initial aberration with RMS values (b) within the range of 0 to 2.236 rad, (c) of 1.0 rad (d) of 2.0 rad. The graphs were averaged from 20 sets of results and the shaded areas marked the standard deviation of the result distribution. The relative correction ability for larger average initial aberration is a consequence of the generally lower initial image intensity for higher aberrations.*

To evaluate practical application on realistic extended specimens, we demonstrated the method on biological samples including fixed Bovine Pulmonary Artery Endothelial (BPAE) cells (FluoCells™ Prepared Slide #1) and 250 $\mu m$ thick fixed Thy1-YFP mouse brain tissue. We selected the 6-foci pattern for this analysis because it had shown the overall best performance on bead images.

For the thin BPAE cells slide experiment, we simulated specimen-induced aberrations by introducing additional aberrations using the same SLM device. Several iterations of correction were run in order to observe convergence. Results are shown in Figure 4 (a-c). For all cases, the algorithm corrected most of the introduced aberration leaving nearly zero remaining aberrations, as can be seen from the insets showing residual wavefront aberration. The algorithm corrected fast for the first two iterations, then maintained a fairly flat response thereafter.

For the thick brain tissue, the aberrations were introduced by the sample's refractive index structure. It was clear that the algorithm converged mostly after the first two iterations and the images are much brighter after correction.

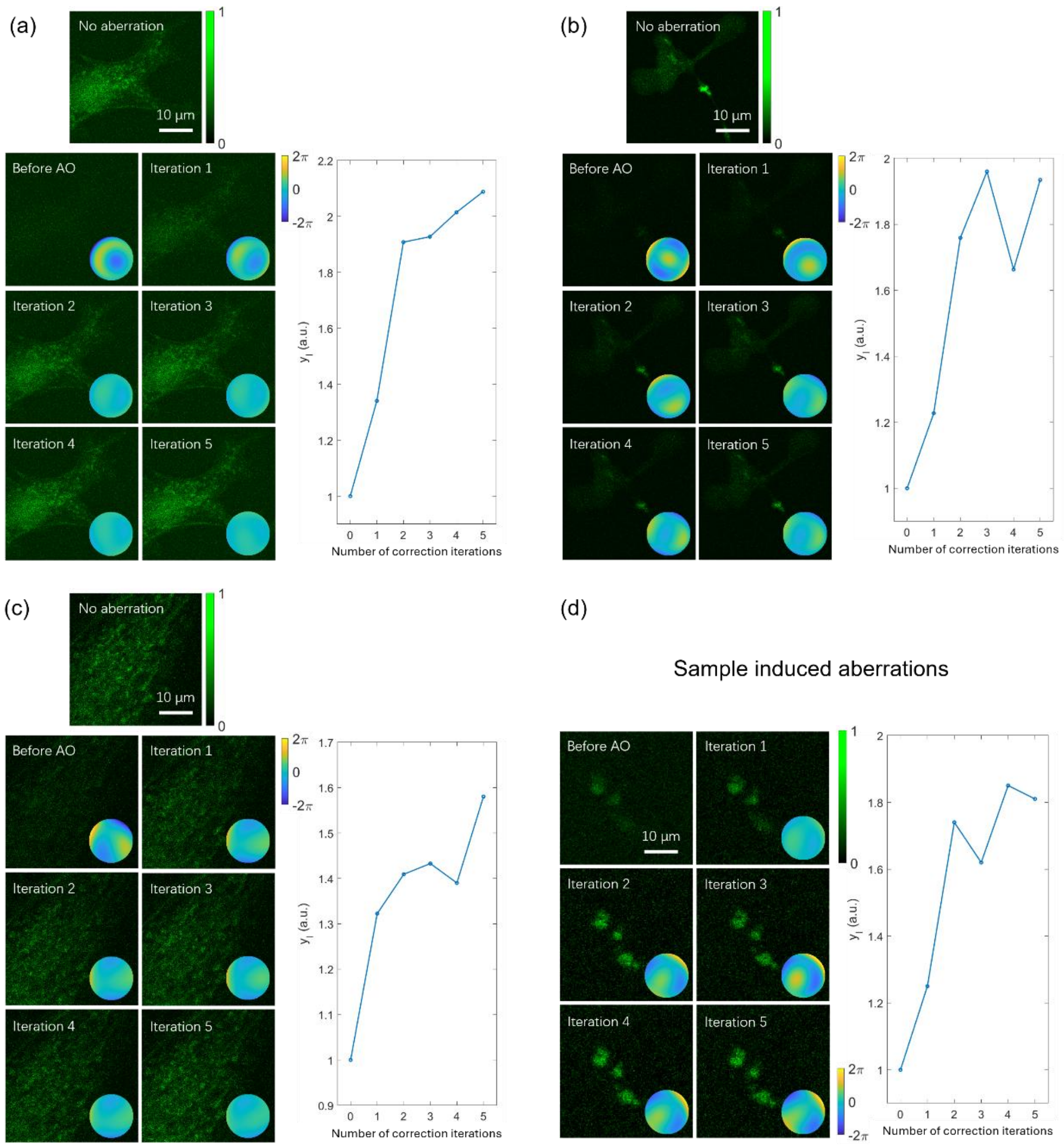

*Figure 4 (a-c) Five iterations of correction on randomly introduced aberrations when imaging microtubules of BPAE cells. Insets showed the introduced aberration and the remaining aberration after each iteration of correction. Graphs showed the variations of the intensity metric as corrections. (d) Five iterations of correction on sample-induced aberrations when imaging neurons in a fixed mouse brain tissue. Insets showed the applied corrections as iterations. The graph showed the intensity metric variation.*

## Conclusion

We have demonstrated that sensorless adaptive optics aberration estimation can be performed from a single image acquisition by using a holographically generated multi-focal array, with each focal spot incorporating a different bias aberration. In contrast to many previous single-image phase retrieval approaches, the proposed method is applicable to imaging extended, non-trivial specimens rather than being restricted to point-like objects.

The effectiveness of aberration estimation depended on the number of foci used. It was observed that more foci did not necessarily lead to better conditioned aberration information, but often showed reduced performance. Similarly, too few focal spots also led to weaker performance, as they provided less phase diversity than more spots, reducing the quality of aberration estimation. In the scope of our work, we found that overall 6 foci worked relatively better than 2 or 10 foci for estimation and correction of 5 modes.

For the cell and tissue specimen demonstrations, two successive estimation–correction iterations were found to achieve a near-converged correction. This behaviour is consistent with the residual estimation errors expected when the initial specimen aberrations are relatively large. Nevertheless, the ability to perform accurate aberration estimation from a single image in each iteration substantially reduces acquisition time compared with conventional multi-frame sensorless AO methods.

The proposed approach therefore demonstrates that spatial multiplexing of bias aberrations, combined with information-theoretic optimisation and neural network-based estimation, provides an effective route towards highly efficient sensorless adaptive optics for two-photon fluorescence microscopy.

## References

[1] M. J. Booth. “Adaptive optical microscopy: the ongoing quest for a perfect image.” Light: Science & Applications 3(4): e165, 2014.

[2] N. Ji. “Adaptive optical fluorescence microscopy.” Nature Methods 14(4): 374-380, 2017.

[3] Q. Zhang et al. “Adaptive optics for optical microscopy.” Biomedical Optics Express 14(4): 1732-1756, 2023.

[4] Q. Hu et al. “A universal framework for microscope sensorless adaptive optics: Generalized aberration representations.” APL Photonics 5(10), 2020.

[5] M. J. Booth et al. “Adaptive aberration correction in a confocal microscope.” Proceedings of the National Academy of Sciences 99(9): 5788-5792, 2002.

[6] P. N. Marsh, D. Burns and J. M. Girkin. “Practical implementation of adaptive optics in multiphoton microscopy.” Optics Express 11(10): 1123-1130, 2003.

[7] D. Débarre et al. “Image-based adaptive optics for two-photon microscopy.” Optics Letters 34(16): 2495-2497, 2009.

[8] A. Facomprez, E. Beaurepaire and D. Débarre. “Accuracy of correction in modal sensorless adaptive optics.” Optics Express 20(3): 2598-2612, 2012.

[9] J. Tang, R. N. Germain and M. Cui. “Superpenetration optical microscopy by iterative multiphoton adaptive compensation technique.” Proceedings of the National Academy of Sciences 109(22): 8434-8439, 2012.

[10] D. Débarre, M. J. Booth and T. Wilson. “Image based adaptive optics through optimisation of low spatial frequencies.” Optics Express 15(13): 8176-8190, 2007.

[11] W. M. Lee and S. H. Yun. “Adaptive aberration correction of GRIN lenses for confocal endomicroscopy.” Optics Letters 36(23): 4608-4610, 2011.

[12] S. W. Paine and J. R. Fienup. “Machine learning for improved image-based wavefront sensing.” Optics Letters 43(6): 1235-1238, 2018.

[13] Y. Nishizaki et al. “Deep learning wavefront sensing.” Optics Express 27(1): 240-251, 2019.

[14] B. P. Cumming and Min Gu. “Direct determination of aberration functions in microscopy by an artificial neural network.” Optics Express 28(10): 14511-14521, 2020.

[15] D. Saha et al. “Practical sensorless aberration estimation for 3D microscopy with deep learning.” Optics Express 28(20): 29044-29053, 2020.

[16] Q. Hu et al. “Universal adaptive optics for microscopy through embedded neural network control.” Light: Science & Applications 12(1): 270, 2023.

[17] M. A. A. Neil, M. J. Booth and T. Wilson. "New modal wave-front sensor: a theoretical analysis." Journal of the Optical Society of America A 17(6): 1098-1107, 2000.

[18] M. A. A. Neil, M. J. Booth and T. Wilson. "Closed-loop aberration correction by use of a modal Zernike wave-front sensor." Optics Letters 25(15): 1083-1085, 2000.

[19] G. P. Andersen et al. “Holographic wavefront sensor.” Optical Engineering 48(8): 085801-085801, 2009.

[20] R. W. Gerchberg and W. O. Saxton. “A practical algorithm for the determination of the phase from image and diffraction plane pictures.” Optik 35(2): 237–246, 1972.

[21] P. Memmolo et al. “Investigation on specific solutions of Gerchberg–Saxton algorithm.” Optics and Lasers in Engineering 52: 206-211, 2014.

[22] B. Zhang, M. J. Booth and Q. Hu. “Information-guided optimization of image-based sensorless adaptive optics methods” Optics Express 34(18): 33074-33086, 2026.

[23] G. Marsaglia. “Choosing a point from the surface of a sphere.” The Annals of Mathematical Statistics 43(2): 645-646, 1972.